\documentclass{ptephy_v1-2009.06525v6}  

\preprintnumber{arXiv:2009.06525}     

\usepackage{amsmath,amssymb,revsymb,graphicx,dcolumn}

\newcommand\ddfracNEW[2]{\displaystyle{\frac{#1}{#2}}}  

\begin{document}

\title{IIB matrix model and regularized big bang}

\author{F.R. Klinkhamer$^\ast$}
\affil{Institute for Theoretical Physics,
Karlsruhe Institute of Technology (KIT),\\
76128 Karlsruhe, Germany \email{frans.klinkhamer@kit.edu}}

\begin{abstract}%
The large-$N$ master field of the Lorentzian IIB matrix model
can, in principle, give rise to a particular degenerate metric
relevant to a regularized big bang. The length parameter of
this degenerate metric is then calculated in terms of the
IIB-matrix-model length scale.
\end{abstract}

\subjectindex{B25,\, B83}


\maketitle

\section{Introduction}
\label{sec:Intro}

Einstein's gravitational field equation~\cite{Einstein1916}
gives, in a cosmological context, the
Friedmann--Lema\^{i}tre--Robertson--Walker (FLRW) 
solution of a homogeneous and isotropic expanding universe with
relativistic matter~\cite{Friedmann1922,Friedmann1924,Lemaitre1927,%
Robertson1935-I,Robertson1936-II,Robertson1936-III,Walker1937}.
This solution has, however, 
a singularity with diverging energy density and curvature: 
the big bang singularity 
at cosmic-time coordinate $t=0$.

Recently, we have suggested another
solution~\cite{Klinkhamer2019-rbb},
which has an additional length parameter $b$.
This solution has maximum values of energy density
and Kretschmann curvature scalar 
proportional to $b^{-2}$ and $b^{-4}$, respectively.
In a way, the length parameter $b$ acts as a
``regulator'' of the big bang singularity and the
new solution has been called the regularized big bang
solution.
This new solution replaces the Friedmann big bang curvature
singularity at $t = 0$ by a ``spacetime defect''
localized at $t = 0$.
The spacetime defect is, in fact, described by a
\emph{degenerate} metric with a vanishing
determinant at $t = 0$. The details of this
new cosmological solution are discussed in
Refs.~\cite{Klinkhamer2020-more,KlinkhamerWang2019-cosm,%
KlinkhamerWang2020-pert} and further information
on this particular type of spacetime defect
appears in Refs.~\cite{KlinkhamerSorba2014,%
Guenther2017,Klinkhamer2019-JPCS}.

Up until now, the length parameter $b$ of the degenerate
metric is a mathematical artifact (regulator).
But it is also possible that $b$ is actually a \emph{remnant}
of a new physics phase that replaces Einstein gravity.
In Appendix~B  of Ref.~\cite{Klinkhamer2020-more} 
and Appendix~C of
Ref.~\cite{KlinkhamerWang2020-pert}, we have explicitly
mentioned loop quantum gravity~\cite{AshtekarSingh2011}
and string theory~\cite{GasperiniVeneziano2002}
as possible candidates for
the physics of this new phase. Especially interesting may be
the nonperturbative formulation of string theory, which
may hold some surprises in store for the nature of the new
phase~\cite{Witten2002}.

A particular formulation of nonperturbative type-IIB superstring
theory (M--theory) is given by the so-called IIB matrix
model~\cite{IKKT-1997,Aoki-etal-review-1999}. As
this model involves only a finite number of matrices
(traceless Hermitian matrices of size $N\times N$,
where $N$ is taken to infinity),
spacetime and gravity must emerge dynamically.
Numerical simulations~\cite{KimNishimuraTsuchiya2012,NishimuraTsuchiya2019}
of the Lorentzian version of
the IIB matrix model suggest, in fact, that a
ten-dimensional classical spacetime
emerges with three ``large'' spatial dimensions behaving differently
from six ``small'' spatial dimensions. The previous
literature~\cite{IKKT-1997,Aoki-etal-review-1999,%
KimNishimuraTsuchiya2012,NishimuraTsuchiya2019} is,
however, not entirely clear on from where precisely the
spacetime points and metric come.

About a year ago,  
we suggested that, in the context of matrix
models, the large-$N$ master field~\cite{Witten1979}
may play a crucial role for the emergence of a classical spacetime.
This suggestion was detailed in Ref.~\cite{Klinkhamer2020-master} and
several toy-model calculations were presented in two follow-up
papers~\cite{Klinkhamer2020-points,Klinkhamer2020-metric}.

We now pose the following question:
does the master field of the Lorentzian IIB matrix model
(assumed to be relevant for the physics of the Universe)
give an emerging spacetime with a particular degenerate metric
that corresponds to the regularized big bang
solution of general relativity?
At this moment, we cannot provide a definite answer, as we do
not know the IIB-matrix-model master field.
However, awaiting the final result on the master field, we can already
investigate what properties the master field would need to have
in order to be able to produce, if at all possible,
an emerging metric resembling the metric of the
regularized big bang solution.
(It is far from obvious that the IIB-matrix-model expression
for the emergent metric can give rise to such a type of metric.)
The present paper is, therefore, solely exploratory in character.

\section{Background material}
\label{sec:Background material}

\subsection{Regularized big bang solution}
\label{subsec:Regularized-big-bang-solution}

In Sec.~\ref{sec:Intro}, we have already mentioned the
main properties of the regularized big bang solution
in general relativity.
Here, we will briefly recall the relevant expressions
of this metric.

The new line element is given
by~\cite{Robertson1935-I,Robertson1936-II,Robertson1936-III,%
Walker1937,Klinkhamer2019-rbb}%
\begin{subequations}\label{eq:reg-bb}
\begin{eqnarray}\label{eq:reg-bb-ds2}
\hspace*{-6mm}
ds^{2}\,\Big|^\text{(RWK)}
&\equiv&
g_{\mu\nu}(x)\, dx^{\mu}\,dx^{\nu} \,\Big|^\text{(RWK)}
=
- \frac{ t^{2}}{b^{2}+ t^{2}}\,d t^{2}
+ a^{2}( t )
\;\delta_{ij}\,dx^{i}\,dx^j\,,
\\[2mm]
\hspace*{-6mm}
b &>& 0\,,
\qquad
a^{2}( t ) > 0\,,
\\[2mm]
\label{eq:ranges-cosmic-coordinates}
\hspace*{-6mm}
 t    &\in& (-\infty,\,\infty)\,,\quad
 x^{i} \in (-\infty,\,\infty)\,,
\end{eqnarray}
\end{subequations}
where we have set $x^{0}=c\, t$ and $c=1$.
The spacetime indices $\mu$, $\nu$ run over $\{0,\, 1,\, 2,\, 3 \}$
and the spatial indices $i $, $j $ over $\{1,\, 2,\, 3 \}$.
Observe that the cosmic-time coordinate $t$ covers the whole of
the real line. The real function $a(t)$ corresponds to
the cosmic scale factor.

The metric from \eqref{eq:reg-bb} is degenerate, having
a vanishing determinant at $ t =0$, and describes
a spacetime defect with a characteristic length scale $b$.
Further references on this type of spacetime defect have
been given in Refs.~\cite{Klinkhamer2019-rbb,Klinkhamer2020-more}.
The similarities and differences 
of the standard Robertson--Walker (RW) metric             
and the degenerate metric \eqref{eq:reg-bb} are discussed     
in a recent review~\cite{Klinkhamer2021-APPB-review}.

Assuming a homogeneous perfect fluid for the matter,
with energy density $\rho_{M}(t)$ and pressure $P_{M}(t)$,
and inserting the metric \eqref{eq:reg-bb}
in the Einstein gravitational field equation
(taking the appropriate limits~\cite{Guenther2017}  for $t=0$)
produces a modified Friedmann equation for $a(t)$,
which has a bounce-type solution with a nonsingular
behavior of the energy density and curvature at $t=0$.
The matter of the homogeneous perfect fluid 
is assumed to satisfy the standard energy conditions, for example the
null energy condition $\rho_{M}+P_{M} \geq 0$.

Specifically, the modified spatially flat Friedmann equation,
the energy-conservation equation of the matter,
and the equation of state of the matter 
are~\cite{Klinkhamer2020-more} 
\begin{subequations}\label{eq:mod-Feqs}
\begin{eqnarray}\label{eq:mod-Feq-1stFeq}
\hspace*{-0mm}&&
\left[1+ \frac{b^{2}}{t^{2}}\,\right]\,
\left( \frac{\dot{a}}{a}\right)^{2}
= \frac{8\pi G}{3}\,\rho_{M}\,,
\\[2mm]
\label{eq:mod-Feq-rhoMprimeeq}
\hspace*{-0mm}&&
\dot{\rho}_{M}+ 3\;\frac{\dot{a}}{a}\;
\Big[\rho_{M}+P_{M}\Big] =0\,,
\\[2mm]
\label{eq:mod-Feq-EOS}
\hspace*{-0mm}&&
P_{M} = P_{M}\big(\rho_{M}\big)\,,
\end{eqnarray}
\end{subequations}
where the overdot stands for differentiation with respect to $t$
and $G$ is Newton's gravitational coupling constant.
For relativistic matter ($P_{M}/\rho_{M}=1/3$),
the regular solution of \eqref{eq:mod-Feqs}
reads~\cite{Friedmann1922,Friedmann1924,Lemaitre1927,%
Robertson1935-I,Robertson1936-II,Robertson1936-III,%
Walker1937,Klinkhamer2019-rbb}
\begin{equation}\label{eq:regularized-Friedmann-asol}
a(t)\,\Big|_\text{(rel.\;mat.)}^\text{(FLRWK)}
\propto
\big(t^{2}+b^{2}\big)^{1/4}\,,
\end{equation}
with $\rho_{M}(t) \propto 1/(t^{2}+b^{2})$.
The new solution \eqref{eq:regularized-Friedmann-asol}
appears, in slightly different notation,
as Eq.~(3.7) in Ref.~\cite{Klinkhamer2019-rbb}.
The review~\cite{Klinkhamer2021-APPB-review}
elaborates on how the degenerate metric \eqref{eq:reg-bb}
with cosmic scale factor $a(t)$
from \eqref{eq:regularized-Friedmann-asol}
satisfies the Hawking--Penrose cosmological singularity
theorems~\cite{HawkingPenrose1970}.
Further discussion of the resulting bouncing cosmology
appears in Refs.~\cite{KlinkhamerWang2019-cosm,%
KlinkhamerWang2020-pert}, but here we highlight only one result.

Consider a perturbative
\textit{Ansatz} around the ``bounce'' at $t=0$,  
\begin{subequations}\label{eq:pert-Ansatz-a-rhoM}
\begin{eqnarray}
a(t) &=& 1+ a_{2}\,\big(t/b\big)^{\,2}+ \cdots\,,
\\[2mm]
\rho_{M}(t) &=& \rho_{M,0}\,\Big(1+ r_{2}\,\big(t/b\big)^{\,2}+ \cdots\Big)\,,
\end{eqnarray}
\end{subequations}
with the energy-density scale $\rho_{M,0}>0$ and
real constants $a_n$ and $r_n$, for $n \geq 2$.
The modified Friedmann
equation \eqref{eq:mod-Feq-1stFeq}
then gives the following parametric relation:
\begin{equation}
\label{eq:relation-GN-rhoM0-b}
1/G  \equiv 1/(l_\text{Planck})^{2}
\sim b^{2}\;\rho_{M,0}\,,
\end{equation}
where we have assumed $a_{2} \sim 1$
and have set $\hbar=1$ and $c=1$ in the
definition of the Planck length \big[recall that, in general,
we have
$(l_\text{Planck})^{2}\equiv \hbar\,G/c^3$\,\big].
From the measured value of $G$,
we get $l_\text{Planck} \approx 1.62 \times 10^{-35}\,
\text{m}$; 
see Chapter~43 of Ref.~\cite{MisnerThorneWheeler2017}  
for further discussion. The relation \eqref{eq:relation-GN-rhoM0-b}
will be used in Sec.~\ref{sec:Conclusion}.

For comparison with later results, we give explicitly 
the degenerate metric (a rank-2 covariant tensor)
corresponding to the line element \eqref{eq:reg-bb},
\begin{eqnarray}
\label{eq:gmunu-reg-bb-metric}
g_{\mu\nu}\,\Big|^\text{(RWK)}  &=&
\begin{cases}
  -\;  \ddfracNEW{\textstyle t^{2}}{\textstyle t^{2}+b^{2}} \,,
  &  \;\;\text{for}\;\;\mu=\nu=0 \,,
 \\[4mm]
 a^{2}(t) \,,
 &  \;\;\text{for}\;\;\mu=\nu=m\in \{1,\, 2,\, 3\} \,,
 \\[4mm]
 0 \,,   &  \;\;\text{otherwise}\,.
 \end{cases}
\end{eqnarray}
The inverse metric (a rank-2 contravariant tensor) is
simply given by the matrix inverse
(cf. p.~201 of Ref.~\cite{MisnerThorneWheeler2017})%
\begin{eqnarray}
\label{eq:gmunu-reg-bb-inverse-metric}
g^{\mu\nu}\,\Big|^\text{(RWK)}  &=&
\begin{cases}
  -\; \ddfracNEW{\textstyle t^{2}+b^{2}}{\textstyle t^{2}} \,,
 &  \;\;\text{for}\;\;\mu=\nu=0 \,,
 \\[4mm]
a^{-2}(t) \,,
 &  \;\;\text{for}\;\;\mu=\nu=m\in \{1,\, 2,\, 3\} \,,
 \\[4mm]
 0 \,,   &  \;\;\text{otherwise}\,,
 \end{cases}
\end{eqnarray}
which, for $b^{2} > 0$, has a divergent $g^{00}$ component at $t=0$.

\subsection{Emergent spacetime metric}
\label{subsec:Emergent-spacetime-metric}

The IIB matrix model is extremely simple to formulate,
having a finite number of matrices,
but extremely hard to evaluate and interpret.
More specifically, the model has
a finite number of \mbox{$N\times N$}
traceless Hermitian matrices ($N$ is taken to infinity later).
Details of the IIB matrix model are given in the original
papers~\cite{IKKT-1997,Aoki-etal-review-1999}
and have been briefly reviewed in 
Ref.~\cite{Klinkhamer2020-master}.
Here, we only recall what is needed for the further discussion.

Adapting Eq.~(4.16) of Ref.~\cite{Aoki-etal-review-1999}
to our master-field approach, we have obtained
the following expression for the emergent inverse
metric~\cite{Klinkhamer2020-master}:
\begin{equation} \label{eq:emergent-inverse-metric-D-spacetime-dim}
g^{\mu\nu}(x) \sim
\int_{\mathbb{R}^{D}} d^{D}y\;
\langle\langle\, \rho(y)  \,\rangle\rangle
\; (y-x)^{\mu}\,(y-x)^{\nu}\;f(y-x)\;r(x,\,y)\,,
\end{equation}
with spacetime dimension $D=10$ for the original matrix model
and continuous spacetime coordinates $x^{\mu}$.
These spacetime coordinates $x^{\mu}$ have the dimension of length,
which traces back to the IIB-matrix-model length scale $\ell$
that has been introduced in the path integral~\cite{Klinkhamer2020-master}.
The average $\langle\langle\, \ldots  \,\rangle\rangle$
in the integrand of \eqref{eq:emergent-inverse-metric-D-spacetime-dim}
will be discussed shortly, after some other explanations have been
given.

We refer to Refs.~\cite{Klinkhamer2020-master,%
Klinkhamer2020-points} for the details on how
the discrete spacetime points $\widehat{x}^{\,\mu}_{k}$, with
index $k\in \{1,\, \ldots \,,K\}$, are extracted
from the bosonic master field $\widehat{\underline{A}}^{\,\mu}$.
This bosonic master field corresponds to a set of ten $N\times N$
traceless Hermitian matrices for
$N=K n$, with positive integers $K$ and $n$.
The limit $K \to \infty$ carries along the limit $N \to \infty$,
provided $n$ stays constant or increases
(the role of $n$ will be explained below).

The quantities  entering the integral
\eqref{eq:emergent-inverse-metric-D-spacetime-dim} are
the density function
\begin{equation} \label{eq:rho-def}
\rho(x) \;\equiv \;
\sum_{k=1}^{K}\;\delta^{(D)} \big(x- \widehat{x}_{k}\big)
\end{equation}
for the emergent spacetime points $\widehat{x}^{\,\mu}_{k}$
as obtained in Refs.~\cite{Klinkhamer2020-master,%
Klinkhamer2020-points}
and the dimensionless density correlation function $r(x,\,y)$
defined by
\begin{equation} \label{eq:r-def}
\langle\langle\,\rho(x)\,\rho(y) \,\rangle\rangle \;\equiv \;
\langle\langle\, \rho(x)\,\rangle\rangle\;
\langle\langle\,\rho(y) \,\rangle\rangle\;r(x,\,y)\,.
\end{equation}
In \eqref{eq:emergent-inverse-metric-D-spacetime-dim},
there is also a localized symmetric real function $f(y-x)$,
which appears in the effective
action~\cite{Aoki-etal-review-1999,Klinkhamer2020-master}
of a low-energy scalar
degree of freedom $\sigma$ hopping over the discrete
spacetime points $\widehat{x}^{\,\mu}_{k}$,
\begin{equation} \label{eq:Seff-sigma}
S_\text{eff}[\sigma] \sim
\sum_{k,\,l}\; \frac{1}{2}\,f\big(\widehat{x}_{k}-\widehat{x}_{l}\big)\;
\big( \sigma_{k}- \sigma_{l}  \big)^{2}\,,
\end{equation}
where $\sigma_{k}$ is the field value
at the point $\widehat{x}_{k}$
(the scalar degree of freedom $\sigma$
arises from a perturbation of the
master field $\underline{\widehat{A}}^{\,\mu}$
and $\sigma$ has the dimension of length;
see Appendix~A of Ref.~\cite{Klinkhamer2020-master}
for a toy-model calculation).
As this function $f(x)=f\big(x^{0},\,  x^{1},\, \ldots \,,\,x^{D-1}\big)$
has the dimension of $1/(\text{length})^{2}$,
the inverse metric $g^{\mu\nu}(x)$
from \eqref{eq:emergent-inverse-metric-D-spacetime-dim}
is manifestly dimensionless.
The metric $g_{\mu\nu}$ is obtained as the matrix inverse
of $g^{\mu\nu}$.

The extraction
procedure~\cite{Klinkhamer2020-master,Klinkhamer2020-points}
of the discrete spacetime points $\widehat{x}^{\,\mu}_{k}$
relies on $n\times n$ blocks positioned adjacently
along the diagonal of the $N\times N$
matrices $\widehat{\underline{A}}^{\,\mu}$ of the
bosonic master field (there are $K$ blocks on each diagonal).
Very briefly, the coordinates of the
$K$ discrete spacetime points $\widehat{x}^{\,\mu}_{k}$
are obtained as follows: the $k$-th $n\times n$ block on the
diagonal of the single $N\times N$ matrix
$\widehat{\underline{A}}^{\,\sigma}$
(where $\sigma$ is a fixed index) gives a
single coordinate $\widehat{x}^{\,\sigma}_{k}$
as the average of the real eigenvalues of this particular $n\times n$ block.
The average $\langle\langle\, \ldots  \,\rangle\rangle$
entering \eqref{eq:emergent-inverse-metric-D-spacetime-dim}
and \eqref{eq:r-def} then corresponds
to averaging over different block sizes $n$
and different block positions along the diagonal in the master field.
The details of this averaging procedure still need to be clarified,
but this does not affect the present discussion.

A few heuristic remarks may help to clarify
expression \eqref{eq:emergent-inverse-metric-D-spacetime-dim}
for the emergent inverse metric.
In the standard continuum theory [i.e., a scalar field
$\sigma(x)$ propagating  over a given continuous spacetime manifold
with metric $g_{\mu\nu}(x)$],
two nearby points $x'$ and $x''$ have approximately equal field values,
$\sigma(x') \sim \sigma(x'')$,
and two distant points $x'$ and $x'''$ generically
have different field values, $\sigma(x') \ne \sigma(x''')$.
The logic is inverted for our discussion.
Two approximately equal field values, $\sigma_1 \sim \sigma_2$,
still have a relatively small action \eqref{eq:Seff-sigma}
if $f(\widehat{x}_1-\widehat{x}_2)\sim 1$, 
and inserting $f\sim 1$ in
\eqref{eq:emergent-inverse-metric-D-spacetime-dim}
gives a ``large'' value for the inverse metric $g^{\mu\nu}$
and, hence, a ``small'' value for the metric $g_{\mu\nu}$,
meaning that the spacetime points $\widehat{x}_1$ and $\widehat{x}_2$
are close  (in units of $\ell$).
Similarly, two very different field values $\sigma_1$ and $\sigma_3$
have a relatively small action (\ref{eq:Seff-sigma})
if $f(\widehat{x}_1-\widehat{x}_3)$ $\sim$ $0$ and
inserting $f\sim 0$ in \eqref{eq:emergent-inverse-metric-D-spacetime-dim}
gives a ``small'' value for the inverse metric $g^{\mu\nu}$
and, hence, a ``large'' value for the metric $g_{\mu\nu}$,
meaning that the spacetime points $\widehat{x}_1$ and $\widehat{x}_3$
are separated by a large distance (in units of $\ell$).

In the following, we will focus on the four ``large'' spacetime
dimensions~\cite{KimNishimuraTsuchiya2012,NishimuraTsuchiya2019}
and we have, for the emergent inverse metric,
\begin{subequations} \label{eq:emergent-inverse-metric-4D}
\begin{eqnarray}
\label{eq:emergent-inverse-metric-4D-gmunu}
g^{\mu\nu}(x) &\sim&
\int_{\mathbb{R}^{4}} d^{4}y\;
\rho_\text{av}(y) \; (y-x)^{\mu}\,(y-x)^{\nu}\;f(y-x)\;r(x,\,y)\,,
\\[2mm]
\label{eq:emergent-inverse-metric-4D-rho-av}
\rho_\text{av}(y) &\equiv& \langle\langle\, \rho(y)  \,\rangle\rangle\,,
\end{eqnarray}
\end{subequations}
with an effective spacetime dimension $D=4$ and
the abbreviated notation $\rho_\text{av}$.
Perhaps it is not even necessary to do this additional
averaging of $\rho$ in the integrand of
\eqref{eq:emergent-inverse-metric-4D-gmunu}, 
as that is already taken care of  
by the $N\to\infty$ limit~\cite{Klinkhamer2020-metric}.

In Ref.~\cite{Klinkhamer2020-metric}, we have
rewritten the integral \eqref{eq:emergent-inverse-metric-4D-gmunu}
somewhat
by using the integration variables $z^{\mu}\equiv y^{\mu}-x^{\mu}$
and introducing new functions $h$ and $\overline{r}$.
The resulting integral and the required definitions are%
\begin{subequations} \label{eq:emergent-inverse-metric-4D-z}
\begin{eqnarray}
\label{eq:emergent-inverse-metric-4D-z-gmunu}
g^{\mu\nu}(x) &\sim&
\int_{\mathbb{R}^{4}} d^{4}z\;
\rho_\text{av}(z+x) \; z^{\mu}\,z^{\nu}\;h(z)\;\overline{r}(x,\,z+x)\,,
\\[2mm]
h(y-x)  &\equiv& f(y-x)\;\widetilde{r}(y-x)\,,
\\[2mm]
r(x,\,y) &\equiv& \widetilde{r}(y-x)\;\overline{r}(x,\,y)\,,
\end{eqnarray}
\end{subequations}
where the new function $\overline{r}(x,\,y)$
has a more complicated dependence on $x$ and $y$
than the combination $x-y$, but the function
is still symmetric, $\overline{r}(x,\,y)=\overline{r}(y,\,x)$.
The advantage of using \eqref{eq:emergent-inverse-metric-4D-z-gmunu}
is that the $x$-dependence in the integrand
has now been insolated in only two functions,
$\rho_\text{av}$ and $\overline{r}$.

For later use, we recall that the action of
the ten-dimensional Lorentzian IIB
matrix model~\cite{IKKT-1997,Aoki-etal-review-1999,%
KimNishimuraTsuchiya2012,NishimuraTsuchiya2019}
contains coupling constants $\widetilde{\eta}_{\mu\nu}$,
for indices $\mu,\,\nu\in \{0,\, 1,\, \ldots\, ,\,9\}$.
Reduced to $D=4$ dimensions, these coupling constants
are given by
\begin{equation} \label{eq:etatilde-munu-4D}
\widetilde{\eta}_{\mu\nu} =
\begin{cases}
 -1 \,,   &  \;\;\text{for}\;\;\mu=\nu=0 \,,
 \\[2mm]
 +1 \,,   &  \;\;\text{for}\;\;\mu=\nu=m\in \{1,\, 2,\, 3\} \,,
 \\[2mm]
 0 \,,   &  \;\;\text{otherwise} \,.
\end{cases}
\end{equation}
We emphasize that the above $16$ numbers are only
coupling constants and not yet a metric.

The purpose of the present paper
is to investigate the
integral \eqref{eq:emergent-inverse-metric-4D-z-gmunu}. It is
not at all obvious that a Lorentzian inverse metric could appear
with the required singular behavior.
Indeed, we want to determine what would be required of the
unknown functions $\rho_\text{av}$, $h$, and $\overline{r}$
[which trace back to the IIB-matrix-model master field],
so that the integral \eqref{eq:emergent-inverse-metric-4D-z-gmunu}
gives the inverse metric \eqref{eq:gmunu-reg-bb-inverse-metric},
which has a divergent $g^{00}$ component at $t=0$.

\section{Emergent degenerate metric}
\label{sec:Emergent-degenerate-metric}

\subsection{Basic idea}
\label{subsec:Basic-idea}

By choosing an appropriate length unit,  we set
the IIB-matrix-model length scale $\ell$ to unity, $\ell=1$.
In this way, the coordinates $\widehat{x}^{\,\mu}_{k}$
of the discrete emerging spacetime points
are effectively dimensionless,
and the same holds for the continuous
spacetime coordinates $x^{\mu}$  used in
Sec.~\ref{subsec:Emergent-spacetime-metric}.
Moreover, we write, in a cosmological context,
these continuous spacetime coordinates as follows:%
\begin{subequations}\label{eq:xmu-cosmological}
\begin{eqnarray}
x^{\mu} &=& \big(x^{0},\,x^{1},\, x^{2},\, x^{3}\big)\,,
\\[2mm]
x^{0} &=& \widetilde{c}\,t =t\,,
\end{eqnarray}
\end{subequations}
where $t$ is interpreted as the cosmic-time coordinate
and $\widetilde{c}$ is set to unity by an appropriate choice
of the time unit. The cosmic-time coordinate $t$
is also effectively dimensionless.

In order to obtain an emergent inverse metric with a possibly divergent
$g^{00}$ component at $t=0$, the convergence properties of the
$z^{0}$ integral in \eqref{eq:emergent-inverse-metric-4D-z-gmunu}
need to be relaxed. Instead of the
factor $\exp\big[-\big(z^{0}\big)^{2}\big]$ in $h(z)$
as used by Ref.~\cite{Klinkhamer2020-metric}
for the standard spatially flat RW 
inverse metric,
we consider the following structure of the
function $h(z)$ entering \eqref{eq:emergent-inverse-metric-4D-z-gmunu}:
\begin{equation}
\label{eq:h-z-heuristics}
h(z) \sim
\frac{1}{\big(z^{0}\big)^{2}+1}\;
\exp\Big[-\big(z^{1}\big)^{2}-\big(z^{2}\big)^{2}-\big(z^{3}\big)^{2}\,\Big]\,.
\end{equation}
Focussing on the $z^{0}$ integral and neglecting other
contributions, we then have
\begin{subequations}\label{eq:g00integral-g11integral-heuristics}
\begin{eqnarray}
g^{00} &\sim&
\int_{-z^{0}_\text{\,cutoff}}^{z^{0}_\text{\,cutoff}}\,dz^{0}\;
\frac{\big(z^{0}\big)^{2}}{\big(z^{0}\big)^{2}+1}\,,
\\[2mm]
g^{11} &\sim&
\int_{-z^{0}_\text{\,cutoff}}^{z^{0}_\text{\,cutoff}}\,dz^{0}\; \frac{1}{\big(z^{0}\big)^{2}+1}\,,
\end{eqnarray}
\end{subequations}
where the first integral diverges linearly 
as $z^{0}_\text{\,cutoff}\to\infty$ but the second does not.  

Next, we must obtain $z^{0}_\text{\,cutoff} \sim 1/t^{2}$
and we use, for that, the following \textit{Ansatz}:
\begin{equation}
\label{eq:rbar-heuristics}
\overline{r}(x,\,z+x) \sim
\frac{p_{1}}{1+\big(z^{0}\big)^{2}\,\big(x^{0}\big)^{4}}
+
\frac{p_{2}}{1+\big(z^{0}\big)^{2}\,\big(z^{0}+x^{0}\big)^{4}}\,,
\end{equation}
where $x^0$ is identified with the cosmic-time coordinate
$t$ and where, later, we set $p_{1}=p_{2}=1/\pi$.
Note that the above function $\overline{r}(x,\,z+x)$,
with equal $p_{1}$ and $p_{2}$, is symmetric
in its arguments $x$ and $z+x$,
\mbox{which explains the appearance of the second term proportional
to $p_{2}$.}

From \eqref{eq:emergent-inverse-metric-4D-z}
with $\rho_\text{av}=1$ and
the \textit{Ans\"{a}tze} \eqref{eq:h-z-heuristics}  
and \eqref{eq:rbar-heuristics},  we find that
the integrals with $p_{1}$ can be done analytically.
The integrals with $p_{2}$ are more complicated but can be dealt with
after a Taylor expansion with respect to $x^0=t$.
The following structure is obtained:
\begin{subequations}\label{eq:absg00result-absg11result-heuristics}
\begin{eqnarray}
\big|g^{00}\big| &\propto& \frac{1}{t^{2}}+\text{O}(1)\,,
\\[2mm]
\big|g^{11}\big| &\sim& \text{O}(1)\,.
\end{eqnarray}
\end{subequations}
Further work is needed to get a $t$-independent term
in $\big|g^{00}\big|$ exactly equal to unity and the Lorentzian signature.
In a first reading, it is possible to skip
the technical details and move forward to Sec.~\ref{sec:Conclusion}.

\subsection{Core structure}
\label{subsec:Core-structure}

With the basic idea of the previous subsection [namely, a mild
cutoff on the $z^{0}$ integral of \eqref{eq:emergent-inverse-metric-4D-z}
at values of order $\pm \,t^{-2}\,$], we have not yet
obtained the core structure of the desired inverse
metric \eqref{eq:gmunu-reg-bb-inverse-metric}.
For that, we need an extended \textit{Ansatz} with additional
freedom carried by four real parameters
$\{\alpha,\,\beta,\, \gamma,\,\delta\}$.
Remark that ``core structure'' refers to the inner structure
of the spacetime defect~\cite{Klinkhamer2019-rbb},
which, in this case, concerns the
time coordinate and corresponds to a divergent $g^{00}$ component.

Specifically, we take the following \textit{Ansatz} functions:
\begin{subequations}\label{eq:Ansatz-rbar-h-rho}
\begin{eqnarray}
\label{eq:Ansatz-rbar}
\hspace*{-7mm}
\overline{r}(x,\,z+x) &=&
\frac{p_{1}}{1+\alpha\,\big(z^{0}\big)^{2}\,\big(x^{0}\big)^{4}}
+
\frac{p_{2}}{1+\alpha\,\big(z^{0}\big)^{2}\,\big(z^{0}+x^{0}\big)^{4}}\,,
\\[2mm]
\label{eq:Ansatz-h}
\hspace*{-7mm}
h(z) &=&
\xi\;\frac{\beta}{1+\gamma\,\big(z^{0}\big)^{2}}\;
\exp\Big[-\big(z^{1}\big)^{2}-\big(z^{2}\big)^{2}-\big(z^{3}\big)^{2}\, \Big]\;
\Big(
\widetilde{\eta}_{00}+
\nonumber\\[1mm]
\hspace*{-7mm}
&&
+\widetilde{\eta}_{11}\,\delta\,\left[\,\zeta\,\big(z^{1}\big)^{2}-1\,\right]
+\widetilde{\eta}_{22}\,\delta\,\left[\,\zeta\,\big(z^{2}\big)^{2}-1\,\right]
+\widetilde{\eta}_{33}\,\delta\,\left[\,\zeta\,\big(z^{3}\big)^{2}-1\,\right]
\Big)\,,
\\[2mm]
\label{eq:Ansatz-rho}
\hspace*{-7mm}
\rho_\text{av}(z+x) &=& 1\,,
\\[2mm]
\label{eq:Ansatz-alpha-and-gamma-positive}
\hspace*{-5mm}
\alpha &>& 0,\, \quad \gamma >0\,,
\end{eqnarray}
\end{subequations}
where we set, as before, $x^0=t$ and $p_{1}=p_{2}=1/\pi$.
One of the constants $\xi$ or $\beta$ in \eqref{eq:Ansatz-h}
is superfluous, but we keep them both in order to ease the comparison with
the previous calculation of Ref.~\cite{Klinkhamer2020-metric}.
The $h(z)$ \textit{Ansatz}  involves, in addition, the coupling constants
$\widetilde{\eta}_{\mu\nu}$ from the Lorentzian IIB matrix model
reduced to $D=4$ dimensions, as given by \eqref{eq:etatilde-munu-4D}.
For a different way of obtaining a Lorentzian signature
in the emergent inverse metric,
see Appendix~B of Ref.~\cite{Klinkhamer2020-master}
and Appendix~D of Ref.~\cite{Klinkhamer2021-APPB-review}.

Inserting the \textit{Ansatz} functions \eqref{eq:Ansatz-rbar-h-rho}
into the emergent-inverse-metric
expression \eqref{eq:emergent-inverse-metric-4D-z},
we can perform all
integrals analytically, except for the $z^{0}$ integral involving
the $p_{2}$ term.
For that integral, we make a Taylor expansion in $x^0=t$
and then integrate analytically the resulting Taylor coefficients.
As explained in Ref.~\cite{Klinkhamer2020-metric},
we set
\begin{subequations}\label{eq:zeta-xi}
\begin{eqnarray}
\zeta &=& 2\,,
\\[2mm]
\xi  &=& 1/\pi^{3/2}\,,
\end{eqnarray}
\end{subequations}
and obtain the following result ($m \in \{1,\, \ldots\, ,\,3\}$
is the spatial index):
\begin{subequations}\label{eq:gmunu-general-structure}
\begin{eqnarray}
g^{00} &\sim& (-1)\;
\Big[ c^{00}_{-2}\,t^{-2}+ c^{00}_{0} +\text{O}(t^{2}) \Big]\,,
\\[2mm]
g^{mm} &\sim& (+1)\;\Big[c^{mm}_{0} +\text{O}(t^{2}) \Big] \,,
\end{eqnarray}
\end{subequations}
with all other components $g^{\mu\nu}$ vanishing
by symmetry [the integrand of \eqref{eq:emergent-inverse-metric-4D-z}
then has a single factor $z^{1}$, $z^{2}$, or $z^{3}\,$].
The coefficients $c^{\mu\nu}_{n}$ in \eqref{eq:gmunu-general-structure}
are functions of the four real \textit{Ansatz} parameters $\{\alpha,\,\beta,\, \gamma,\,\delta\}$.

In order to simplify the discussion, we immediately fix
\begin{equation}
\label{eq:alphabeta-bar}
\overline{\alpha}=1\,, \quad \overline{\beta}=10^{2}\,,
\end{equation}
so that we only need to determine the appropriate values of
the parameters $\gamma$ and $\delta$.
In fact, the $c^{00}_{0}$ coefficient now only depends on
the parameter $\gamma$,
as $\delta$ is absent and $\alpha$ and $\beta$ have been fixed to
the numerical values  \eqref{eq:alphabeta-bar}.
From the requirement
\begin{equation}
\label{eq:c-00-0}
\widetilde{c}^{\;00}_{0}=1\,,
\end{equation}
where the tilde indicates the use of \eqref{eq:alphabeta-bar},
we obtain a seventh-order algebraic equation for $\sqrt{\gamma}$,
which has two positive real roots.
The analytic expressions for these two roots are rather cumbersome
and we will just give their numerical values,
\begin{subequations}\label{eq:gamma-1-2}
\begin{eqnarray}
\gamma_{1} &\approx& 10.1337\,,
\\[2mm]
\gamma_{2} &\approx& 34.5392\,.
\end{eqnarray}
\end{subequations}
For definiteness, we take the first root from \eqref{eq:gamma-1-2}
and set
\begin{equation} \label{eq:gamma-bar}
\overline{\gamma}=\gamma_{1} \approx 10.1337\,.
\end{equation}
Having found a suitable value for $\gamma$, we turn to
the resulting
coefficient $c^{mm}_{0}$ of the inverse-metric component $g^{mm}$.
From the requirement
\begin{equation}
\label{eq:c-mm-0}
\widetilde{c}^{\;mm}_{0}=1\,,
\end{equation}
where the tilde indicates the use of \eqref{eq:alphabeta-bar}
and \eqref{eq:gamma-bar},
we obtain a linear equation for $\delta$
and find the following solution:
\begin{equation} \label{eq:delta-bar}
\overline{\delta}\approx 0.517689\,.
\end{equation}

To summarize, we have, from the \textit{Ansatz} functions \eqref{eq:Ansatz-rbar-h-rho} and the parameters
\begin{equation} \label{eq:alphabetagammadelta-bar}
\big\{\overline{\alpha},\,\overline{\beta},\,
\overline{\gamma},\,\overline{\delta}\big\}
=
\big\{1,\,10^{2},\,10.1337,\,0.517689\big\}\,,
\end{equation}
the following result for the emergent inverse metric
as given by the expression \eqref{eq:emergent-inverse-metric-4D-z}:
\begin{equation}
\label{eq:gmunu-core-structure}
\hspace*{-5mm}
g^{\mu\nu}\,\Big|^\text{(core-structure)} \sim
\begin{cases}
 (-1)\;\Big[ \big(\overline{\beta}/\overline{\gamma}\big)\,t^{-2}+
            1 +\text{O}(t^{2}) \Big]
 \,,   &  \;\;\text{for}\;\;\mu=\nu=0 \,,
 \\[2mm]
 (+1)\;\Big[ 1 +\text{O}(t^{2}) \Big] \,,
 &  \;\;\text{for}\;\;\mu=\nu=m\in \{1,\, 2,\, 3\} \,,
 \\[2mm]
 0 \,,   &  \;\;\text{otherwise}\,,
 \end{cases}
\end{equation}
where the numerical value
of $\overline{\beta}/\overline{\gamma}$ is of the order of 
$10$ (the actual numerical value will be given shortly).

Comparing to the general-relativity inverse
metric \eqref{eq:gmunu-reg-bb-inverse-metric},
we interpret the first two nontrivial terms of $g^{00}$
from \eqref{eq:gmunu-core-structure} as follows:
\begin{subequations}\label{eq:g00-core-structure}
\begin{eqnarray}\label{eq:g00-core-structure-metric}
g^{00}\,\Big|^\text{(core-structure)} &\sim& (-1)\;
\left[\frac{ b^{2}/\ell^{2} + t^{2}/\ell^{2} }{ t^{2}/\ell^{2} }
+ \ldots \,\right]\,,
\\[2mm]
\label{eq:g00-core-structure-b2-over-ell2}
b^{2}/\ell^{2} &=& \overline{\beta}/\overline{\gamma}\,,
\end{eqnarray}
\end{subequations}
where $\ell$ is the length scale of the IIB matrix model
that we have previously set to unity.
With the parameter values \eqref{eq:alphabetagammadelta-bar}, we have
\begin{equation} \label{eq:b2-result}
b^{2}/\ell^{2}=\overline{\beta}/\overline{\gamma}\approx 9.8681\,,
\end{equation}
but different numerical values are obtained if, for example,
the $\beta$ value is changed away from the value $10^{2}$ or if
the root $\gamma_{2}$ is chosen instead of $\gamma_{1}$.
The general parametric behavior of the $t^{-2}$ coefficient in
$g^{00}$ follows by adapting the elementary integral for $g^{00}$
in Sec.~\ref{subsec:Basic-idea} and gives
\begin{equation}
\label{eq:b2-parametric-behavior}
b^{2}
\sim
\frac{\beta}{\sqrt{\alpha}\,\gamma}\;\ell^{2}\,,
\end{equation}
for the particular \textit{Ansatz} \eqref{eq:Ansatz-rbar-h-rho}.

\subsection{First approximation}
\label{subsec:First-approximation}

In the previous subsection, we have shown that, in principle, the emergent
inverse-metric expression \eqref{eq:emergent-inverse-metric-4D-z}
can give the core structure 
of the inverse metric \eqref{eq:gmunu-reg-bb-inverse-metric},
with an explicit
numerical value of the classical-gravity length parameter $b$
in units of the IIB-matrix-model length scale $\ell$.
See, in particular,
the results \eqref{eq:g00-core-structure}, \eqref{eq:b2-result},
and \eqref{eq:b2-parametric-behavior}.

We now want to check that the higher-order terms
in $t$ of \eqref{eq:g00-core-structure-metric} can be made to vanish.
For that, we will use, instead of \eqref{eq:Ansatz-rho},
a nontrivial \textit{Ansatz} of the $\rho_\text{av}$
function. Specifically, we take
\begin{equation}
\label{eq:Ansatz-rho-with-rn-terms}
\rho_\text{av}(z+x)= 1+
\left(  \sum_{k=0}^{2}\,r_{2k}\;\left(z^{0}+x^0\right)^{2k} \right)\;
\exp\Big[-\left(z^{0}+x^0\right)^{2}\Big] \,,
\end{equation}
with real parameters $r_{n}$  and an explicit exponential factor
to guarantee the convergence of the $z^{0}$ integral
\big(the $r_{n}$ terms will, for this reason, not modify the
coefficient of the $t^{-2}$ term in $g^{00}$\big).
Keeping the $\gamma$ parameter equal to the numerical value
$\overline{\gamma}$  from \eqref{eq:gamma-bar}
but allowing for a change in the numerical value of $\delta$,
we find that the coefficients $c^{\mu\nu}_{n}$
of \eqref{eq:gmunu-general-structure} have the following dependence:
\begin{subequations}\label{eq:coeff-cmunun-dependence}
\begin{eqnarray}
\overline{c}^{\,00}_{0} &=&
\overline{c}^{\,00}_{0}\big(r_{0},\, r_{2},\, r_{4}\big)\,,
\\[2mm]
\overline{c}^{\,00}_{2} &=&
\overline{c}^{\,00}_{2}\big( r_{0},\, r_{2},\, r_{4} \big)\,,
\\[2mm]
\overline{c}^{\,11}_{0} &=&
\overline{c}^{\,11}_{0}\big( r_{0},\, r_{2},\, r_{4},\,\delta \big)\,,
\end{eqnarray}
\end{subequations}
where the overbar indicates the use of the numerical
values \eqref{eq:alphabeta-bar} and \eqref{eq:gamma-bar}.

Demanding
\begin{equation}
\label{eq:coeff-cmunun-conditions}
\overline{c}^{\,00}_{0}=1\,, \quad
\overline{c}^{\,00}_{2}=0\,, \quad
\overline{c}^{\,11}_{0}=1\,,
\end{equation}
gives three algebraic equations for the three parameters
$\{ r_{0},\, r_{2},\, \delta\}$ with the following solutions: 
\begin{subequations}\label{eq:r0r2delta-bar}
\begin{eqnarray}
\overline{r}_{0} &\approx&
-0.297254 - 0.305679\,r_{4}\,,
\\[2mm]
\overline{r}_{2} &\approx&
+0.491944 - 0.850548\,r_{4}\,,
\\[2mm]
\overline{\delta} &\approx&
\frac{12.013 - 3.77531\,r_{4}}{23.026 - 7.55061\,r_{4}}\,,
\end{eqnarray}
\end{subequations}
which are still functions of the free parameter $r_{4}$. 
[Note that $r_{0}=r_{2}=r_{4}=0$ is not a
solution of the conditions \eqref{eq:coeff-cmunun-conditions}.] 
The corresponding inverse metric reads
\begin{subequations}\label{eq:gmunu-first-approx}
\begin{eqnarray}
\label{eq:gmunu-first-approx-metric}
\hspace*{-10mm}
g^{\mu\nu}\,\Big|^\text{(first approx.)} &\sim&
\begin{cases}
 (-1)\;\Big[ \big(\overline{\beta}/\overline{\gamma}\big)\,t^{-2}+
            1 +\text{O}(t^{4}) \Big]
 \,,   &  \;\;\text{for}\;\;\mu=\nu=0 \,,
 \\[2mm]
 (+1)\;\Big[ 1 + \overline{c}_{2}\,t^{2} + \text{O}(t^{4}) \Big] \,,
 &  \;\;\text{for}\;\;\mu=\nu=m\in \{1,\, 2,\, 3\} \,,
 \\[2mm]
 0 \,,   &  \;\;\text{otherwise}\,,
 \end{cases}
\\[2mm]
\label{eq:gmunu-first-approx-betabar-over-gammabar}
\hspace*{-10mm}
\overline{\beta}/\overline{\gamma} &\approx& 9.8681\,,
\\[2mm]
\label{eq:gmunu-first-approx-cbar2}
\hspace*{-10mm}
\overline{c}_{2} &\approx&
\frac{4.17896 + 0.862081\,r_{4}}{23.026 - 7.55061\,r_{4}}\,,
\end{eqnarray}
\end{subequations}
which is a significant improvement compared to the
core-structure result \eqref{eq:gmunu-core-structure}.
The result \eqref{eq:gmunu-first-approx}
corresponds, in fact, to a first approximation of the desired
inverse metric valid to order $t^{4}$.

We can invert the map \eqref{eq:gmunu-first-approx-cbar2} and
obtain the required input value $r_{4}$
for a desired value of $c_{2}\,$,
\begin{equation}
\label{eq:r4input}
r_{4,\,\text{input}} \approx
-\frac{4.84753 - 26.7098\,c_{2,\,\text{desired}}}
  {1 + 8.75859\,c_{2,\,\text{desired}}}\,.
\end{equation}
In this way, we can get \emph{any}
Taylor coefficient $c_{2}$ in the $g^{mm}$ component
from \eqref{eq:gmunu-first-approx} by choosing an appropriate value
of the \textit{Ansatz} parameter $r_{4}$.

From \eqref{eq:gmunu-first-approx-metric},
we obtain by matrix inversion the
diagonal metric $g_{\mu\nu}$ which has the following $00$ component:
\begin{equation}
\label{eq:g00-first-approx-metric}
g_{00}\,\Big|^\text{(first approx.)} \sim
 (-1)\;
\frac{t^{2}}{\overline{\beta}/\overline{\gamma}+t^{2}+\text{O}(t^{6})}\,.
\end{equation}
It is already clear that this metric is degenerate,
with a vanishing determinant at $t=0$, but we postpone
further discussion of this point to the next subsection.

\subsection{Conjectured final result}
\label{subsec:Conjectured-final-result}

As indicated on the left-hand side of
\eqref{eq:gmunu-first-approx-metric},
we consider that result to be a first approximation 
of the inverse metric \eqref{eq:gmunu-reg-bb-inverse-metric}, 
as derived from the IIB-matrix-model
master field under the assumptions stated.
Better approximations, with more and more Taylor coefficients
for $g^{mm}$ and more and more $t^n$ terms vanishing in $g^{00}$,
will follow from higher orders in the
\textit{Ansatz} function $\rho_\text{av}$ from \eqref{eq:Ansatz-rho-with-rn-terms}
and possible further extensions of the
\textit{Ansatz} functions $\overline{r}$ and $h$.
This procedure has been tested in Ref.~\cite{Klinkhamer2020-metric}
for the standard spatially flat RW inverse metric.

The final result for the emergent inverse metric is expected
to have the following structure (in units with $\ell=1$):
\begin{eqnarray}
\label{eq:gmunu-conjectured-inverse-metric}
\hspace*{-8mm}
g^{\mu\nu}\,\Big|^\text{(final-result)}
 &\stackrel{?}{\sim}&
\begin{cases}
  -\ddfracNEW{\textstyle t^{2}+c_{-2}}{\textstyle t^{2}}
 \,,   &  \;\;\text{for}\;\;\mu=\nu=0 \,,
 \\[3mm]
  1 + c_{2}\,t^{2} + c_{4}\,t^{4} + \dots  \,,
 &  \;\;\text{for}\;\;\mu=\nu=m\in \{1,\, 2,\, 3\} \,,
 \\[3mm]
 0 \,,   &  \;\;\text{otherwise}\,,
 \end{cases}
\end{eqnarray}
where the question mark indicates
that, strictly speaking, this is a conjectured result.
The real dimensionless coefficients $c_{n}$ in $g^{mm}$     
of \eqref{eq:gmunu-conjectured-inverse-metric} result from 
the requirement that $t^{2n}$ terms, for integer $n>0$, 
vanish in $g^{00}$. The emergent metric is given by the matrix inverse
of \eqref{eq:gmunu-conjectured-inverse-metric},%
\begin{eqnarray}
\label{eq:gmunu-conjectured-metric}
\hspace*{-8mm}
g_{\mu\nu}\,\Big|^\text{(final-result)}
&\stackrel{?}{\sim}&
\begin{cases}
   - \ddfracNEW{\textstyle t^{2}}{\textstyle t^{2}+c_{-2}}
 \,,   &  \;\;\text{for}\;\;\mu=\nu=0 \,,
 \\[3mm]
 \ddfracNEW{\textstyle 1}{\textstyle 1 + c_{2}\,t^{2} + c_{4}\,t^{4} + \dots} \,,
 &  \;\;\text{for}\;\;\mu=\nu=m\in \{1,\, 2,\, 3\} \,,
 \\[3mm]
 0 \,,   &  \;\;\text{otherwise}\,,
 \end{cases}
\end{eqnarray}
which has, for $c_{-2}> 0$, a vanishing determinant at $t=0$.
In short, the emergent metric \eqref{eq:gmunu-conjectured-metric},
obtained from the expression \eqref{eq:emergent-inverse-metric-4D-z}
with  appropriate \textit{Ansatz} functions and parameters,
is degenerate.

The emergent metric \eqref{eq:gmunu-conjectured-metric}
has indeed the structure of the metric \eqref{eq:gmunu-reg-bb-metric}, 
with the following effective parameters:
\begin{subequations}\label{eq:b2-eff-a-eff}
\begin{eqnarray}
\label{eq:b2-eff}
b^{2}_\text{eff} &\sim& c_{-2}\;\ell^{2}\,,
\\[2mm]
\label{eq:a-eff}
a^{2}_\text{eff}(t) &\sim& 1-c_{2}\;\big(t/\ell\big)^{2}
+ \text{O}(t^{4}/\ell^{4})\,,
\end{eqnarray}
\end{subequations}
where the IIB-matrix-model length scale $\ell$ has been
restored and where we omit the question marks as we have explicit
results for the coefficients shown.
Indeed, the leading coefficients are given by
$c_{-2} \approx \overline{\beta}/\overline{\gamma}>0$
from \eqref{eq:b2-result}
and $c_{2} \approx \overline{c}_{2}(r_4)$
from \eqref{eq:gmunu-first-approx-cbar2},
for the particular \textit{Ansatz} functions \eqref{eq:Ansatz-rbar},
\eqref{eq:Ansatz-h}, and \eqref{eq:Ansatz-rho-with-rn-terms}
and \textit{Ansatz} parameters \eqref{eq:alphabeta-bar}
and \eqref{eq:gamma-bar}.
If the \textit{Ansatz} parameter $r_4$
in \eqref{eq:gmunu-first-approx-cbar2}
is chosen appropriately, we get $c_{2} \approx \overline{c}_{2} <0$ in
the square of the cosmic scale factor \eqref{eq:a-eff},
so that the emergent classical spacetime corresponds to
the spacetime of a nonsingular cosmic bounce at $t=0$,
as obtained in Refs.~\cite{Klinkhamer2019-rbb,Klinkhamer2020-more}
from Einstein's gravitational field equation.
The proper cosmological interpretation of the emergent classical
spacetime will be discussed further in Sec.~\ref{sec:Conclusion}.

\section{Conclusion}
\label{sec:Conclusion}

In the present article, we have started an exploratory
investigation of how a new physics phase can
give an emerging classical spacetime with an effective metric
where the big bang singularity has been tamed~\cite{Klinkhamer2019-rbb}.

In order to be explicit, we have used the
IIB matrix model~\cite{IKKT-1997,Aoki-etal-review-1999},
which has been suggested as a nonperturbative definition of
type-IIB superstring theory. If we interpret the numerical
results~\cite{KimNishimuraTsuchiya2012,NishimuraTsuchiya2019}
from the Lorentzian IIB matrix model as corresponding to an approximation
of the genuine master field~\cite{Witten1979},
then it appears that spacetime
points emerge with three ``large'' spatial dimensions and
six ``small'' spatial dimensions. But the numerical simulations
are still far removed from providing results on the required
density and correlation functions that build the inverse
metric~\cite{Aoki-etal-review-1999,Klinkhamer2020-master}.

For the moment, we have adopted a leapfrogging strategy by
jumping over the actual analytic or numeric evaluation of the
IIB-matrix-model master field and by simply assuming certain types
of behavior of the density and correlation functions that enter
the inverse-metric
expression \eqref{eq:emergent-inverse-metric-D-spacetime-dim}.
The explicit  goal of the present article is to establish
what type of functions are required
in \eqref{eq:emergent-inverse-metric-D-spacetime-dim}
to get, if at all possible, 
an inverse metric with the behavior 
shown in \eqref{eq:gmunu-reg-bb-inverse-metric}.
[Note that, in principle, the origin of the
expression \eqref{eq:emergent-inverse-metric-D-spacetime-dim}
need not be the IIB matrix model but can be an entirely
different theory, as long as the emerging inverse metric
is given by a multiple integral with the same basic structure.]

For the integral \eqref{eq:emergent-inverse-metric-D-spacetime-dim},
we have indeed been able to find
suitable functions (these functions are, most likely, not unique),
which give an emerging classical spacetime with an effective metric
where the big bang singularity has been tamed.
In fact, the big bang singularity is effectively
regularized by a nonzero length parameter $b_\text{eff}$ that
is now calculated in terms of the
IIB-matrix-model length scale $\ell \,$; see
the last paragraph of Sec.~\ref{subsec:Conjectured-final-result}.
One important lesson, from the comparison with our
previous calculation~\cite{Klinkhamer2020-metric}
of the Minkowski and RW metrics,
appears to be that the relevant correlation functions
must have long-range tails in the time direction,
in order to get a divergent behavior
of $g^{00}$, as explained in Sec.~\ref{subsec:Basic-idea}.

Note that we have not yet obtained the
effective (Einstein?) gravitational field equation
and the corresponding solution of the metric.
Instead, we have used a general constructive expression
for the inverse metric, as given
by \eqref{eq:emergent-inverse-metric-4D-z} after some redefinitions.
The further consistency of the emerging field theories
may then restrict the values of some of the parameters
entering our explicit \textit{Ansatz} functions
\eqref{eq:Ansatz-rbar},
\eqref{eq:Ansatz-h}, and \eqref{eq:Ansatz-rho-with-rn-terms},
fixing, for example, the values of $\alpha$ and $\beta$,
or even demanding different functional forms of the functions
$\rho_\text{av}(z+x)$, $h(z)$, and $\overline{r}(x,\,z+x)$.

Expanding on the previous paragraph, we observe that
the IIB matrix model not only produces a classical spacetime
but also its matter content~\cite{Aoki-etal-review-1999}.
Now, the IIB matrix model in the formulation of
Ref.~\cite{Klinkhamer2020-master} has a single length scale $\ell$,
so that, for the cosmological quantities \eqref{eq:pert-Ansatz-a-rhoM}
near the bounce at $t=0$, we expect an energy-density
scale $\rho_{M,0} \sim 1/\ell^{4}$.
If, moreover, general covariance~\cite{Aoki-etal-review-1999}
and the Einstein gravitational field equation are recovered,
we have from the relation \eqref{eq:relation-GN-rhoM0-b}
with $b \sim \ell$ the following parametric relation:%
\begin{equation}\label{eq:l-Planck-eff}
l_\text{Planck,\,eff}  \stackrel{?}{\sim} \ell\,,
\end{equation}
where $l_\text{Planck,\,eff}$ corresponds
to $\sqrt{G_\text{eff}}$
(using units to set $\hbar_\text{eff}$ and $c_\text{eff}$
to unity) and where the question mark indicates that this
is a conjectured result.
If correct, the emergent Planck length would,
not surprisingly, be of the same order as the
IIB-matrix-model length scale $\ell$. Reading
\eqref{eq:l-Planck-eff} from right to left and inserting
the experimental values for $G$, $\hbar$, and $c$ on the left-hand
side, we would also have an estimate for the actual value of the
unknown IIB-matrix-model length scale $\ell$,
\begin{equation}\label{eq:l-numerical-value}
\ell \stackrel{?}{\sim} 1.62 \times 10^{-35}\,\text{m}\,,
\end{equation}
where the ``experimental'' numerical value for the Planck length  
was already given a few lines below \eqref{eq:relation-GN-rhoM0-b}.

The cosmological interpretation of
the emergent classical spacetime is perhaps as follows.
The new physics phase is assumed to be described by the IIB matrix model
and the corresponding large-$N$ master field gives rise
to the points and the metric of a classical spacetime.
If the master field has an appropriate structure,
the emergent metric has a tamed big bang, with a metric similar to 
the degenerate metric \eqref{eq:gmunu-reg-bb-metric} of general relativity, 
but now having an effective length parameter $b_\text{eff}$
proportional to the IIB-matrix-model length scale $\ell \,$.
In fact, one possible interpretation is that the new physics phase
has produced a universe-antiuniverse pair~\cite{BoyleFinnTurok2018},
that is, a  ``universe'' for $t>0$ and an ``antiuniverse'' for $t<0$.

As a final comment on our main result $b_\text{eff} \sim \ell$
from \eqref{eq:b2-eff} and the conjectured
result \eqref{eq:l-Planck-eff},
we recall that we have used a IIB-matrix-model length scale $\ell$
that was introduced directly into the
path integral~\cite{Klinkhamer2020-master}.
But a more subtle origin of the length scale $\ell$
is certainly not excluded. One example of such an origin would be,
in the emerging massless relativistic quantum field theory
from the matrix model, the appearance of a length scale by the phenomenon
of dimensional transmutation~\cite{ColemanWeinberg1973}.
In any case, assuming the IIB matrix model to be relevant for physics,
progress on fundamental questions such as the origin
of the length scale or the birth of the Universe will only
happen if more is known about the IIB-matrix-model master field.


\end{document}